\documentclass[12pt]{article}

\usepackage[a4paper,text={16.3cm,22cm}]{geometry}
\usepackage{amsmath,amsfonts,braket,amssymb,bm,bbm,xcolor,graphicx}
\usepackage[small,labelfont=bf]{caption}
\usepackage{cite}
\usepackage[
    bookmarksnumbered=true,
    urlcolor=blue,
    linkbordercolor=red,
    citebordercolor=green,
    bookmarksopen=true
    ]{hyperref}

\allowdisplaybreaks
\numberwithin{equation}{section}

\newcommand{\spac}{{\hspace{0.3mm}}}

\def\lsim{\mathrel{\rlap{\lower4pt\hbox{\hskip1pt$\sim$}}
     \raise1pt\hbox{$<$}}}         
\def\gsim{\mathrel{\rlap{\lower4pt\hbox{\hskip1pt$\sim$}}
     \raise1pt\hbox{$>$}}}         

\title{$B_s\to K^0\bar K^0$ beyond the Standard Model}
  
\begin{document}

\begin{titlepage}

\begin{flushright}
{\small
MITP-25-013
}
\end{flushright}

\makeatletter
\vskip0.8cm
\pdfbookmark[0]{\@title}{title}
\begin{center}
{\Large \bf\boldmath \@title}
\end{center}
\makeatother

\vspace{0.5cm}
\begin{center}
Yuval Grossman,$^a$ Matthias Neubert,$^{a,b}$ Yosef Nir,$^c$ Yogev Shpilman,$^c$\\ and Yehonatan Viernik$^c$ \\[6mm]
\textsl{${}^a$ Department of Physics \& LEPP, Cornell University, Ithaca, NY 14853, USA\\[0.3cm]
${}^b$PRISMA$^+$ Cluster of Excellence \& Mainz Institute for Theoretical Physics\\
Johannes Gutenberg University, Staudingerweg 9, 55128 Mainz, Germany\\[0.3cm]
${}^c$ Department of Particle Physics and Astrophysics, Weizmann Institute of Science\\ 
Rehovot 7610001, Israel}
\end{center}

\vspace{0.6cm}
\pdfbookmark[1]{Abstract}{abstract}
\begin{abstract}
Within the Standard Model, the branching fraction of the rare decay $B_s\to K^0\bar K^0$ is related to other decay rates and CP asymmetries through the approximate $SU(3)$ flavor symmetry of the strong interactions and the heavy-quark limit. Three such relations were shown to be violated at a level of about $3\sigma$ each. By means of a systematic search for new-physics explanations of these puzzles, we find that possible solutions are highly fine-tuned and either excluded by other data or rather implausible. The tight correlation between $B_s\to K\bar K$ and $B\to\pi K$ decays, which is maintained even in the presence of flavor-specific new-physics operators, plays a central role in our analysis.
\end{abstract}

\vfill\noindent\rule{0.4\columnwidth}{0.4pt}\\
\hspace*{2ex} {\small \textit{E-mail:} \href{mailto:yg73@cornell.edu}{yg73@cornell.edu}, \href{mailto:matthias.neubert@uni-mainz.de}{matthias.neubert@uni-mainz.de}, \href{mailto:yosef.nir@weizmann.ac.il}{yosef.nir@weizmann.ac.il}, \\
\hspace*{2ex} \phantom{E-mail: } \href{yogev.shpilman@weizmann.ac.il}{yogev.shpilman@weizmann.ac.il}, \href{mailto:yehonatan.viernik@weizmann.ac.il}{yehonatan.viernik@weizmann.ac.il}}

\end{titlepage}

{\hypersetup{hidelinks}
\pdfbookmark[1]{Contents}{ToC}
\setcounter{tocdepth}{2}
\tableofcontents}

\section{Introduction}

Precision studies of rare weak decays of $B$ mesons can provide powerful tests of the Standard Model (SM). Deviations between measured decay rates and CP asymmetries and the corresponding SM predictions may hint at new-physics (NP) effects at scales significantly above the electroweak scale. The CP-averaged branching fraction ${\rm Br}(B_s\to K^0\bar K ^0$) was measured by the LHCb \cite{LHCb:2020wrt} and BELLE \cite{Belle:2015gho} experiments, with the result
\begin{equation}
   {\rm Br}(B_s\to K^0\bar{K}^0) = (1.76\pm0.31)\cdot 10^{-5} \,.
\end{equation}
To investigate the implications of this measurement, it is convenient to define three ratios of CP-averaged decay rates, defined as \cite{Amhis:2022hpm}
\begin{equation}\label{eq:defR}
\begin{aligned}
   R_{KK}^{sd} &\equiv \left| \frac{V_{td}}{V_{ts}} \right|^2 
    \frac{\Gamma(B_s\to K^0\bar K^0)}{\Gamma(B_d\to K^0\bar K^0)} \,, \\
   R_{KK}^{ss} &\equiv \frac{\Gamma(B_s\to K^0\bar K^0)}{\Gamma(B_s\to K^+ K^-)} \,, \\
   R_{\pi K}^{ud} &\equiv \frac{\Gamma(B^+\to\pi^+ K^0)}{\Gamma(B_d\to\pi^- K^+)} \,,
\end{aligned}
\end{equation}
along with the time-dependent CP asymmetries in $B_s$ decays into final CP eigenstates \cite{ParticleDataGroup:2024cfk},
\begin{equation}
   {\cal A}^s_f(t)
   \equiv \frac{\Gamma_{\bar B_s\to f}(t) - \Gamma_{B_s\to f}(t)}%
               {\Gamma_{\bar B_s\to f}(t) + \Gamma_{B_s\to f}(t)}
   = \frac{-C_f^s \cos(\Delta m_s\spac t) + S_f^s\sin(\Delta m_s\spac t)}%
          {\cosh(\Delta\Gamma_s\spac t/2) - A^{\Delta\Gamma}_f\sinh(\Delta\Gamma_s\spac t/2)} \,,
\end{equation}
where $\Delta m_s$ and $\Delta\Gamma_s$ are the mass and width splitting between the two $B_s$ mass eigenstates. 

The two decay modes in the first ratio are related to each other by exchanging all down and strange quarks, which also changes the relevant CKM parameters. On the other hand, the two decay modes in the second and third ratio are related by exchange of up and down quarks and involve the same CKM parameters. We will see in Section~\ref{sec:QCDF} that theoretical predictions for these ratios are far more accurate than those for the individual decay rates. In the SM, using the $SU(3)$ flavor symmetry of the strong interactions, and neglecting subleading decay topologies, one obtains the predictions \cite{Amhis:2022hpm}
\begin{equation}\label{eq:RssRud}
\begin{aligned}
   \big( R_{KK}^{sd} \big)_{\rm SM} &\simeq 1 \,, \\
   \big( R_{KK}^{ss} - S_{K^+ K^-}^s \cot\gamma \big)_{\rm SM} &\simeq 1 \,, \\
   \big( R_{KK}^{ss}/R_{\pi K}^{ud} \big)_{\rm SM} &\simeq 1 \,,
\end{aligned}
\end{equation}
where $\gamma$ is one of the angles of the unitarity triangle. The experimental values of the three ratios are \cite{ParticleDataGroup:2024cfk}
\begin{equation}\label{eq:expR}
\begin{aligned}
   R_{KK}^{sd} &= 0.62\pm 0.14 \,, \\
   R_{KK}^{ss} &= 0.65\pm 0.13 \,, \\
   R_{\pi K}^{ud} &= 1.11\pm 0.04 \,,
\end{aligned}
\end{equation}
and along with the experimental value $S_{K^+ K^-}^s=0.14\pm 0.05$ this yields
\begin{equation}\label{eq:expsd-ud}
\begin{aligned}
   R_{KK}^{ss} - S_{K^+ K^-}^s \cot\gamma &= 0.58\pm 0.13 \,, \\
   R_{KK}^{ss}/R_{\pi K}^{ud} &= 0.58\pm 0.13 \,.
\end{aligned}
\end{equation}
Thus, each of the predictions in \eqref{eq:RssRud} is violated at about the $3\sigma$ level.

In deriving the SM predictions, contributions from $SU(3)$ flavor-symmetry breaking~\cite{Gronau:1995hm}, electroweak penguins \cite{Gronau:1995hn}, and final-state rescattering \cite{Gronau:2012gs} were neglected. The leading ``factorizable'' sources of $SU(3)$ breaking to the ratio $R_{KK}^{sd}$ arise from phase-space and form-factor effects and are given by
\begin{equation}
   R_{KK}^{sd}|_{\rm fact}
   = \frac{m_{B_d}}{m_{B_s}}\,
    \frac{\sqrt{1-\frac{4\spac m_{K^0}^2}{m_{B_s}^2}}}{\sqrt{1-\frac{4\spac m_{K^0}^2}{m_{B_d}^2}}}\,
    \frac{\left( m_{B_s}^2 - m_{K^0}^2 \right)^2}{\left( m_{B_d}^2 - m_{K^0}^2 \right)^2}\,
    \left| \frac{F_0^{B_s\to K}(m_{K^0}^2)}{F_0^{B_d\to K}(m_{K^0}^2)} \right|^2
   \approx 1.077\pm 0.167 \,, 
\end{equation}
where the ratio of the form factors squared equals $1.024\pm 0.159$ \cite{Parrott:2022rgu,Khodjamirian:2017fxg}. This effect is sizable and increases the tension with experiment. However, it was argued in \cite{Amhis:2022hpm} that rescattering effects could also be significant and lower the value of $R_{KK}^{sd}$ so as to make it compatible with experiment. On the other hand, none of the effects that were neglected can make the other two relations in \eqref{eq:expsd-ud} consistent with experiment. Note, in particular, that the constituent quark transition for the decays in the numerators and denominators of the ratios $R_{KK}^{ss}$ and $R_{\pi K}^{ud}$ are the same: $\bar b\to\bar s d\bar d$ for $B_s\to K^0\bar K^0$ and $B^+\to\pi^+K^0$, and $\bar b\to\bar s u\bar u$ for $B_s\to K^+K^-$ and $B_d\to\pi^- K^+$. Thus, assuming isospin symmetry, any violation of the equality $R_{KK}^{ss}=R_{\pi K}^{ud}$ must necessarily result from non-spectator contributions, for which the light quark in the $B$ meson participates in the effective weak vertex (``weak annihilation''). Unless the experimental values of the two observables in \eqref{eq:expsd-ud} change by significant amounts with future measurements, for example by an upward shift of ${\rm Br}(B_s\to K^0\bar K^0)$ by about $3\sigma$, these puzzles call for the presence of sizable NP contributions. The aim of our analysis is to find which NP operators might account for the deviations of the experimental results from the SM predictions, and if a NP explanation of the data is at all plausible.

Various observables in the decays $B_s\to K^{*0}\bar K^{*0}$, $B_s\to K^{0}\bar K^{*0}$, and $B_s\to K^{*0}\bar K^{0}$ also show deviations from the predictions of the SM and the $SU(3)$ flavor symmetry. For recent discussions of these, see \cite{Alguero:2020xca,Li:2022mtc,Biswas:2023pyw,Lizana:2023kei,Berthiaume:2023kmp,Yu:2024vlz}. The $B_s\to K^0\bar K^0$ decay as a probe of NP was discussed in \cite{Ciuchini:2007hx,Descotes-Genon:2011rgs,Bhattacharya:2012hh}, with emphasis on the CP asymmetry in this decay. Predictions for the branching fraction and CP asymmetries in $B_s\to K^0\bar K^0$ were made using QCD factorization \cite{Beneke:2003zv,Wang:2014mua,Bobeth:2014rra} and a global flavor $SU(3)$ fit \cite{Cheng:2014rfa}. A fit to various $B^+$, $B_d$ and $B_s$ decays into two pseudoscalar mesons, including $B_s\to K^0\bar K^0$, was carried out in \cite{Berthiaume:2023kmp}, yielding a $3.6\sigma$ deviation from the SM predictions. None of these studies considered the correlated impact of NP in $B_s\to K\bar K$ and $B\to\pi K$ decays, which plays a central role in our approach.

The plan of this paper is as follows. In Section~\ref{sec:bascon}, we present a first general discussion of the structure of the three ratios using parameterizations in terms of topological amplitudes and argue that a NP explanation necessarily requires isospin-violating non-spectator contributions to the decay amplitudes. In Section~\ref{sec:NPbasis}, we introduce a rather general basis of flavor-specific NP operators extending the operators present in the SM and work out their contributions to the decay amplitudes of interest. Section~\ref{sec:QCDF} contains our predictions obtained using the QCD factorization approach \cite{Beneke:1999br,Beneke:2000ry,Beneke:2001ev} for the SM, for NP models in which the Wilson coefficients of the SM basis operators receive NP effects, and for our more general extension featuring flavor-specific four-fermion operators. We conclude in Section~\ref{sec:con}.

\section{Basic considerations}
\label{sec:bascon}

Based on the structure of the low-energy effective weak Hamiltonian of the SM \cite{Buchalla:1995vs}, and using the general parameterizations introduced in \cite{Beneke:2003zv}, the amplitudes for the five decay rates appearing in the ratios $R_{KK}^{sd}$, $R_{KK}^{ss}$, and $R_{\pi K}^{ud}$ in (\ref{eq:defR}) can be written as 
\begin{align}\label{eq:amplitudes}
   {\cal A}(\bar B_d\to K^0\bar K^0)
   &= \frac{G_F}{\sqrt2}\,\lambda_p^d\,\bigg[ \left( \alpha_4^p - \frac12\spac\alpha_{4,{\rm EW}}^p \right) A_{\bar K K}
    + \left( b_3^p + b_4^p - \frac12\spac b_{3,{\rm EW}}^p - \frac12\spac b_{4,{\rm EW}}^p \right) B_{\bar K K} \notag\\
   &\hspace{2cm} + \left( b_4^p - \frac12\spac b_{4,{\rm EW}}^p \right) B_{K\bar K} \bigg] \,, \notag\\
   {\cal A}(\bar B_s\to K^0\bar K^0)
   &= \frac{G_F}{\sqrt2}\,\lambda_p^s\,\bigg[ \left( \alpha_4^p - \frac12\spac\alpha_{4,{\rm EW}}^p \right) A_{K\bar K}
    + \left( b_3^p + b_4^p - \frac12\spac b_{3,{\rm EW}}^p - \frac12\spac b_{4,{\rm EW}}^p \right) B_{K\bar K} \notag\\
   &\hspace{2cm} + \left( b_4^p - \frac12\spac b_{4,{\rm EW}}^p \right) B_{\bar K K} \bigg] \,, \notag\\
   {\cal A}(\bar B_s\to K^+ K^-)
   &= \frac{G_F}{\sqrt2}\,\lambda_p^s\spac\bigg[\! \left( \delta_{pu}\spac\alpha_1+\alpha_4^p+\alpha_{4,{\rm EW}}^p \right)\! A_{K\bar K} 
    + \!\left(\! b_3^p + b_4^p - \frac12\spac b_{3,{\rm EW}}^p - \frac12\spac b_{4,{\rm EW}}^p \!\right)\! B_{K\bar K} \notag\\
   &\hspace{2cm} + \left( \delta_{pu}\spac b_1 + b_4^p + \spac b_{4,{\rm EW}}^p \right) B_{\bar K K} \bigg] \,,\notag\\
   {\cal A}(B^-\to \pi^-\bar K^0)
   &= \frac{G_F}{\sqrt2}\,\lambda_p^s  \left[ \left( \alpha_4^p - \frac12\spac\alpha_{4,{\rm EW}}^p \right) A_{\pi\bar K}
    + \left( \delta_{pu}\spac b_2 + b_3^p + b_{3,{\rm EW}}^p \right) B_{\pi\bar K} \right] ,\notag\\
   {\cal A}(\bar B_d\to \pi^+ K^-)
   &= \frac{G_F}{\sqrt2}\,\lambda_p^s \left[ \left( \delta_{pu}\spac\alpha_1+\alpha_4^p+\alpha_{4,{\rm EW}}^p \right) A_{\pi\bar K}
    + \left( b_3^p-\frac12\spac b_{3,{\rm EW}}^p \right) B_{\pi\bar K} \right] ,
\end{align}
where $\lambda_p^D\equiv V_{pb} V_{pD}^*$ with $D=d$ or $s$, and a sum over $p=u,c$ is understood. The corresponding amplitudes for the decays of $B_q$ mesons are obtained by replacing $\lambda_p^D\to(\lambda_p^D)^*$. There is a one-to-one correspondence of the $\alpha_i$ and $b_i$ parameters with the topological amplitudes introduced in \cite{Zeppenfeld:1980ex,Gronau:1995hm,Gronau:1995hn} and employed in \cite{Amhis:2022hpm}. Specifically, the tree amplitude $T$ corresponds to $\alpha_1\spac A_{M_1 M_2}$ (with $M_1$ and $M_2$ denoting the final-state mesons), the penguin amplitude $P$ corresponds to $(\alpha_4^c\spac A_{M_1 M_2}+b_3^c\spac B_{M_1 M_2})$, the color-suppressed electroweak penguin amplitude $P_{\rm EW}^C$ corresponds to $\alpha_{4,{\rm EW}}^c\spac A_{M_1 M_2}$, the exchange amplitude $E$ corresponds to $b_1\spac B_{M_1 M_2}$, the annihilation amplitude $A$ corresponds to $b_2\spac B_{M_1 M_2}$, and the penguin annihilation amplitude $P_A$ corresponds to $b_4^c\spac(B_{M_1 M_2}+B_{M_2 M_1})$. Differences between the penguin amplitudes with up- and charm-quarks propagating in the loops are included in the ``rescattering contributions'' in the topological amplitude approach, and the electroweak penguin annihilation amplitudes $b_{3,4,{\rm EW}}^p$ are usually neglected in this framework.  

We stress that the different topological amplitudes depend on the nature of the participating mesons. More accurately, one should write $\alpha_i\equiv\alpha_i(\bar B_q\to M_1 M_2)$ and $b_i\equiv b_i(\bar B_q\to M_1 M_2)$, where the order $M_1$ and $M_2$ is given by the order of the arguments on the quantities
\begin{equation}\label{eq:AandBdef}
\begin{aligned}
   A_{M_1 M_2} &= i\,m_{B_q}^2\spac F_0^{B_q\to M_1}(0)\spac f_{M_2} \,, \\
   B_{M_1 M_2} &= i\spac f_{B_q}\spac f_{M_1}\spac f_{M_2} \,,
\end{aligned}
\end{equation}
which depend on meson transition form factors and decay constants.\footnote{Compared with \cite{Beneke:2003zv}, we have extracted a factor $G_F/\sqrt2$ from these definitions.} 
In the QCD factorization approach discussed in Section~\ref{sec:QCDF}, one finds that the dependence of the parameters $\alpha_i$ and $b_i$ on the mesons participating in the decay only enters at $\mathcal{O}(\alpha_s)$ and solely via their light-cone distribution amplitudes (LCDAs). This dependence is therefore rather mild. The weak annihilation amplitudes are power suppressed $\sim\Lambda_{\rm QCD}/m_b$ in the heavy-quark limit \cite{Beneke:1999br,Beneke:2000ry,Beneke:2001ev} and governed by the ratios (with scalings $f_{B_q}\sim m_b^{-1/2}$ and $F_0^{B_q\to M}(0)\sim m_b^{-3/2}$)
\begin{equation}\label{eq:BKKoverAKK}
\begin{aligned}
   \frac{B_{\pi K}}{A_{\pi K}} 
   &= \frac{f_{B_q}\spac f_\pi}{m_{B_q}^2\spac F_0^{B\to\pi}(0)} 
    \approx 3.2\cdot 10^{-3} \,, \\
   \frac{B_{KK}}{A_{KK}} 
   &= \frac{f_{B_q}\spac f_K}{m_{B_q}^2\spac F_0^{B_q\to K}(0)} 
    \approx 3.7\cdot 10^{-3} \,,
\end{aligned}
\end{equation}
where there is no need to distinguish between kaons and anti-kaons. The fact that these ratios are small plays an important role in our considerations.

Within the SM, the contribution of the penguin amplitude $\alpha_4^c$ dominates the $|\Delta S|=1$ decay amplitudes in \eqref{eq:amplitudes}, since it is CKM enhanced compared with the tree amplitude $\alpha_1$ and the penguin amplitude $\alpha_4^u$. The electroweak penguin amplitudes are suppressed by a factor $\alpha/\alpha_s$ relative to the QCD penguin amplitudes. The contribution of $\alpha_4^c$ is the same for the two rates that appear in $R_{KK}^{ss}$, as well as for the two rates that enter $R_{\pi K}^{ud}$, up to isospin-breaking effects. It is the same for all four rates in the limit of unbroken $SU(3)$ flavor symmetry. Neglecting all other contributions would give $R_{KK}^{ss}=1$ and $R_{\pi K}^{ud}=1$. The contribution of the tree amplitude $\alpha_1$ is the next-to-leading effect in the processes $\bar B_s\to K^+ K^-$ and $\bar B_d\to\pi^+ K^-$, which are mediated by the quark transition $\bar b\to \bar uu\bar s$. Since $|\alpha_1|\approx 1$ and $|\alpha_4^c|\approx 0.1$, the ratio $|(\lambda^s_u\spac\alpha_1)/(\lambda^s_c\spac\alpha_4^c)|$ is of ${\cal O}(0.1)$ and leads to deviations of the two ratios from unity at this order. In fact, $S^s_{K^+K^-}\cot\gamma$ measures this deviation, as reflected in (\ref{eq:expsd-ud}) \cite{Amhis:2022hpm}. The experimental value of the direct CP asymmetry $A_{\rm CP}(\bar B_d\to\pi^+ K^-)=-0.0834\pm0.0032$ \cite{ParticleDataGroup:2024cfk}, which results from the interference of $\lambda^s_u\spac\alpha_1$ and $\lambda^s_c\spac\alpha_4^c$, gives further support to this estimate. Due to the strong suppression of electroweak penguin and annihilation amplitudes in the SM, it is impossible to explain the central values of the ratios $R_{KK}^{ss}$ and $R_{\pi K}^{ud}$ without invoking NP effects.

Our aim in this paper is to find possible NP contributions that can account for the experimental values of the three ratios in \eqref{eq:expR} and the related observables in \eqref{eq:expsd-ud}. For the latter two quantities, it is apparent from \eqref{eq:amplitudes} that this can only happen via {\em isospin-violating non-spectator} contributions to the decay amplitudes, parameterized by $b_1$, $b_2$ and $b_{3,{\rm EW}}^p$, $b_{4,{\rm EW}}^p$. We will show in Section~\ref{sec:4.2} that it is possible to reproduce the experimental values of $R_{KK}^{ss}$ and $R_{\pi K}^{ud}$ by modifications of some Wilson coefficients of operators in the effective weak Hamiltonian of the SM, but only at the expense of very large modifications -- much larger than the SM values of the corresponding coefficients -- and strong fine-tuning. This motivates us to consider a more general implementation of NP effects, in which the SM operator basis is extended.

\section{Including flavor-specific NP contributions}
\label{sec:NPbasis}

The explicit form of the effective weak Hamiltonian for flavor-changing neutral current (FCNC) transitions with $|\Delta B|=|\Delta D|=1$ in the SM (with $D=s$ or $d$) can be found in \cite{Buchalla:1995vs}.\footnote{Following \cite{Beneke:1999br}, we exchange the definitions of $Q_1$ and $Q_2$ compared with this reference.} 
In order to search for a NP  explanation of the data, we extend this Hamiltonian by allowing for the presence of flavor-specific NP operators, in such a way that 
\begin{equation}\label{eq:HeffNP}
   \mathcal{H}_{\rm eff}^{b\to D} 
   = \mathcal{H}_{\rm eff}^{b\to D,\,\rm SM}
    + \frac{1}{\Lambda_{\rm NP}^2} \sum_{i=3}^6 \sum_q\,C_{i,q}^D\,Q_{i,q}^D 
    + \mbox{h.c.} \,,
\end{equation}
where the sum over $q$ extends over all light quark flavors $q\ne t$. We define the NP operators 
\begin{equation}\label{eq:NPbasis}
\begin{aligned}
   Q_{3,q}^D 
   &= \left( \bar D_i\spac b_i \right)_{V-A} \left( \bar q_j\spac q_j \right)_{V-A} \,, \\
   Q_{4,q}^D 
   &= \left( \bar D_i\spac b_j \right)_{V-A} \left( \bar q_j\spac q_i \right)_{V-A} \,, \\
   Q_{5,q}^D 
   &= \left( \bar D_i\spac b_i \right)_{V-A} \left( \bar q_j\spac q_j \right)_{V+A} \,, \\
   Q_{6,q}^D 
   &= \left( \bar D_i\spac b_j \right)_{V-A} \left( \bar q_j\spac q_i \right)_{V+A} \,, 
\end{aligned}
\end{equation}
in analogy with the QCD penguin operators of the SM, but with {\rm flavor-specific\/} Wilson coefficients for each quark flavor $q$. A sum over color indices $i,j$ is implied. This is quite a general basis for the purposes of our problem. Flipping $(V\mp A)\to(V\pm A)$ in these operators would yield the same contributions to the $B$-meson decay amplitudes into two pseudoscalar mesons up to an overall sign.\footnote{This is no longer true for decays with vector mesons in the final state.} 
We do not consider the presence of four-quark operators containing scalar and pseudoscalar fermion bilinears, since they are only present in rather baroque extensions of the SM. 

The parameter $\Lambda_{\rm NP}$ denotes a generic NP scale. The $|\Delta B|=|\Delta S|=1$ transitions in the SM are dominated by the penguin contributions, whose magnitude is determined by 
\begin{equation}\label{eq:SMscale}
   \frac{G_F}{\sqrt2}\,|V_{cb} V_{cs}^*|\,|\alpha_4^c(K\bar K)| \approx \frac{1}{(5.6\,\text{TeV})^2} \,.
\end{equation}
Since we do not include any CKM matrix elements in the Wilson coefficients of the NP operators (thus allowing for a generic flavor structure of these contributions), we will choose the normalization such that $\Lambda_{\rm NP}=10$\,TeV, so that for ${\cal O}(1)$ Wilson coefficients the NP effects would be roughly commensurate with the SM. The CP-odd phases of the NP Wilson coefficients $C_{i,q}^D$ are defined with respect to the standard phase conventional for the CKM matrix, so that {\em relative\/} phases between these coefficients and the CKM parameters $V_{pb} V_{pD}^*$ (with $p=u,c$) are physical. The terms shown explicitly in \eqref{eq:HeffNP} describe the decays of $\bar B$ mesons (containing a $b$ quark). The decays of $B$ mesons involve the complex conjugated coefficients $\big(C_{i,q}^D\big)^*$.

The QCD and electroweak penguin operators of the SM can be expressed as 
\begin{equation}\label{SMpenguins}
\begin{aligned}
   Q_i &= \sum_q\,Q_{i,q}^D \,; \quad i=3,4,5,6 \,, \\
   Q_i &= \sum_q\,\frac32\spac e_q\,Q_{\hat i,q}^D \,; \quad i=7,8,9,10\,, \quad \hat i=5,6,3,4 \,,
\end{aligned}
\end{equation}
where $e_q$ denote the electric charges of the quarks in units of $e$. Allowing for NP contributions to {\em individual\/} flavor operators in the basis \eqref{eq:NPbasis} offers a much more general way of introducing NP than changing the Wilson coefficients of one or more of the operators in the effective weak Hamiltonian of the SM. For the special cases $q=D$ and $q=b$, the operators $Q_{3,q}^D$ and $Q_{4,q}^D$ are Fierz equivalent, so one of the operators in each pair is redundant. We prefer to keep both operators in the basis, since different NP models might naturally give rise to one or the other operator. Naively, one would conclude that only the combinations $(C_{3,D}^D+C_{4,D}^D)$ and $(C_{3,b}^D+C_{4,b}^D)$ are physically meaningful. However, it is well known that Fierz relations can be broken in dimensional regularization, and indeed this happens for the penguin contributions arising at one-loop order (see the discussion in the paragraph following equation (23) in \cite{Beneke:2001ev}).

The NP operators in \eqref{eq:HeffNP} give rise to new tree, penguin, and annihilation contributions. Following \cite{Beneke:2003zv}, their contributions to the two-body decay amplitudes of $\bar B$ mesons can be parameterized in terms of flavor operators $\mathcal{T}_A^{\rm NP}$ and $\mathcal{T}_B^{\rm NP}$. We define
\begin{equation}
   \langle M_1 M_2|\,\mathcal{H}_{\rm eff}^{b\to D}\,|\bar B\rangle
   = \frac{G_F}{\sqrt2}\,\sum_{p=u,c} \lambda_p^D\,
    \langle M_1 M_2|\,\mathcal{T}_A^p+\mathcal{T}_B^p\,|\bar B\rangle
    + \frac{1}{\Lambda_{\rm NP}^2}\,
    \langle M_1 M_2|\,\mathcal{T}_A^{\rm NP}+\mathcal{T}_B^{\rm NP}\,|\bar B\rangle \,.
\end{equation}
Explicit expressions for the SM flavor operators $\mathcal{T}_A^p$ and $\mathcal{T}_B^p$ have been presented in equations~(9) and (18) of \cite{Beneke:2003zv}. Analyzing the flavor flow of the relevant Feynman graphs for the different topological amplitudes, we find that the NP contributions to the tree and penguin topologies give rise to the $A$-operators
\begin{equation}\label{A-type}
\begin{aligned}
   \text{tree:} & \qquad A\big([\bar q_s\spac q]\,[\bar q\spac D]\big) \,, \quad 
    A\big([\bar q_s D]\,[\bar q\spac q]\big) \,, \\ 
   \text{penguin:} & \qquad \sum_{q'}\,A\big([\bar q_s\spac q']\,[\bar q' D]\big) \,,
\end{aligned}
\end{equation}
where the quark flavors shown in the two brackets $[\dots]$ refer to the mesons $M_1$ and $M_2$, and $\bar q_s$ denotes the light spectator quark inside the $\bar B$ mesons. The NP contributions to the annihilation amplitudes lead to the $B$-operators
\begin{equation}\label{B-type}
   \text{annihilation:} \qquad \sum_{q'}\,B\big([\bar q\spac q']\,[\bar q' q]\,[\bar D\spac b]\big) \,,
    \quad \sum_{q'}\,B\big([\bar q\spac q']\,[\bar q' D]\,[\bar q\spac b]\big) \,,
\end{equation}
where the three brackets $[\dots]$ refer to the mesons $M_1$, $M_2$, and $\bar B$. We are interested in light final-state mesons, so that the flavors $q$ and $q'$ can be restricted to $u$, $d$, $s$. Only the penguin contributions involve NP operators with flavor labels $q=c,b$. We thus obtain
\begin{equation}\label{eq:TNP}
\begin{aligned}
   \mathcal{T}_A^{\rm NP}
   &= \sum_{q=u,d,s} \bigg[\spac \alpha_{1,q}^{D,\spac{\rm NP}}(M_1 M_2)\,A\big([\bar q_s\spac q]\,[\bar q\spac D]\big)
    + \alpha_{2,q}^{D,\spac{\rm NP}}(M_1 M_2)\,A\big([\bar q_s D]\,[\bar q\spac q]\big) \bigg] \\
   &\quad + \alpha_4^{D,\spac{\rm NP}}(M_1 M_2) \sum_{q'=u,d,s} A\big([\bar q_s\spac q']\,[\bar q'\spac\! D]\big) \,, \\
   \mathcal{T}_B^{\rm NP}
   &= \sum_{q,q'=u,d,s} \bigg[\spac b_{1,q}^{D,\spac{\rm NP}}(M_1 M_2)\,B\big([\bar q\spac q']\,[\bar q' q]\,[\bar D\spac b]\big) 
   + b_{2,q}^{D,\spac{\rm NP}}(M_1 M_2)\,B\big([\bar q\spac q']\,[\bar q'\! D]\,[\bar q\spac b]\big) \bigg] \,.
\end{aligned}
\end{equation}
The hadronic matrix elements of these operators can be evaluated using the relations
\begin{equation}
\begin{aligned}
   \langle M_1 M_2|\,A\big([\dots]\,[\dots]\big)\,|\bar B_{q_s}\rangle
   &= c\spac A_{M_1 M_2} \,, \\
   \langle M_1 M_2|\,B\big([\dots]\,[\dots]\,[\dots]\big)\,|\bar B_{q_s}\rangle
   &= c\spac B_{M_1 M_2} \,,
\end{aligned}
\end{equation}
where the quantities $A_{M_1 M_2}$ and $B_{M_1 M_2}$ have been defined in \eqref{eq:AandBdef}. The constant $c$ is a product of factors of 1, $\pm 1/\sqrt2$ etc.\ from the flavor composition of the mesons $M_1$ and $M_2$. The five decay processes in \eqref{eq:amplitudes}, which are of primary interest to us, do not involve mesons containing identical quark flavors, and one has $c=1$. The parameters $\alpha_{2,q}^{{\rm NP},D}$ do not enter the amplitudes of these decays. We find the NP contributions
\begin{align}\label{eq:NPamplitudes}
   {\cal A}(\bar B_d\to K^0\bar K^0)_{\rm NP}
   &= \frac{1}{\Lambda_{\rm NP}^2} \left[ \left( \alpha_{1,s}^{d,\spac{\rm NP}} + \alpha_4^{d,\spac{\rm NP}} \right) A_{\bar K K}
    + b_{1,s}^{d,\spac{\rm NP}}\spac B_{K\bar K} + \left( b_{1,d}^{d,\spac{\rm NP}} + b_{2,d}^{d,\spac{\rm NP}} \right) B_{\bar K K} \right] , \notag\\
   {\cal A}(\bar B_s\to K^0\bar K^0)_{\rm NP}
   &= \frac{1}{\Lambda_{\rm NP}^2} \left[ \left( \alpha_{1,d}^{s,\spac{\rm NP}} + \alpha_4^{s,\spac{\rm NP}} \right) A_{K\bar K}
    + b_{1,d}^{s,\spac{\rm NP}}\spac B_{\bar K K} + \left( b_{1,s}^{s,\spac{\rm NP}} + b_{2,s}^{s,\spac{\rm NP}} \right) B_{K\bar K} \right] , \notag\\
   {\cal A}(\bar B_s\to K^+ K^-)_{\rm NP}
   &= \frac{1}{\Lambda_{\rm NP}^2} \left[ \left( \alpha_{1,u}^{s,\spac{\rm NP}} + \alpha_4^{s,\spac{\rm NP}} \right) A_{K\bar K}
    + b_{1,u}^{s,\spac{\rm NP}}\spac B_{\bar K K} + \left( b_{1,s}^{s,\spac{\rm NP}} + b_{2,s}^{s,\spac{\rm NP}} \right) B_{K\bar K} \right] , \notag\\
   {\cal A}(B^-\to \pi^-\bar K^0)_{\rm NP}
   &= \frac{1}{\Lambda_{\rm NP}^2} \left[ \left( \alpha_{1,d}^{s,\spac{\rm NP}} + \alpha_4^{s,\spac{\rm NP}} \right) A_{\pi\bar K}
    + b_{2,u}^{s,\spac{\rm NP}}\spac B_{\pi\bar K} \right] , \notag\\
   {\cal A}(\bar B_d\to \pi^+ K^-)_{\rm NP}
   &= \frac{1}{\Lambda_{\rm NP}^2} \left[ \left( \alpha_{1,u}^{s,\spac{\rm NP}} + \alpha_4^{s,\spac{\rm NP}} \right) A_{\pi\bar K}
    + b_{2,d}^{s,\spac{\rm NP}}\spac B_{\pi\bar K} \right] .
\end{align}
The NP penguin coefficients $\alpha_4^{D,\spac{\rm NP}}$ enter in the same way in all cases. Its interference with the dominant penguin amplitude $\alpha_4^c$ cancels out in the ratios $R_{KK}^{ss}$ and $R_{\pi K}^{ud}$ to good approximation. To a lesser extent, up to $SU(3)$ breaking effects this is also true for the ratio $R_{KK}^{sd}$. The remaining flavor-specific coefficients break isospin and $U$-spin symmetry explicitly and can in principle affect the three ratios in different ways.

\section{Predictions in the QCD factorization approach}
\label{sec:QCDF}

The QCD factorization approach developed in \cite{Beneke:1999br,Beneke:2000ry,Beneke:2001ev}, which was applied to a large class of two-body $B$-meson decays into pseudoscalar and vector mesons in \cite{Beneke:2003zv}, provides a systematic framework for analyzing hadronic two-body decays of $B$ mesons in the heavy-quark limit. It allows one to study the effects of $SU(3)$ symmetry breaking, electroweak penguins, weak annihilation, and final-state rescattering in a model-independent way, up to corrections suppressed by $\Lambda_{\rm QCD}/m_b$. For the input parameters, we use the values compiled by the Particle Data Group \cite{ParticleDataGroup:2024cfk} where available. For the $\bar B_q\to M$ heavy-to-light transition form factors, the meson decay constants, the Gegenbauer moments of the meson LCDAs, and the first inverse moments of the $B$-meson LCDAs, we use the values collected in Table~3 of \cite{Biswas:2023pyw}. We work at leading power in $\Lambda_{\rm QCD}/m_b$ but include the leading ``chirally-enhanced'' power corrections proportional to the ratios
\begin{equation}\label{eq:rchidef}
   r_\chi^\pi = \frac{2\spac m_\pi^2}{m_b(\mu)\,2\spac m_q(\mu)} \,, \qquad
   r_\chi^K = \frac{2\spac m_K^2}{m_b(\mu)\,(m_q+m_s)(\mu)} \,,
\end{equation}
with $m_q\equiv\frac12(m_u+m_d)$, in a heuristic way, following \cite{Beneke:2000ry,Beneke:2001ev}. Note that these quantities are formally of $\mathcal{O}(\Lambda_{\rm QCD}/m_b)$ but numerically of $\mathcal{O}(1)$. Beyond the leading power, one encounters some endpoint-divergent convolution integrals over light-mesons LCDAs. As suggested in \cite{Beneke:2001ev}, we parameterize these divergences by means of quantities $X_H$ (for hard spectator scattering) and $X_A$ (for weak annihilation), for which we assume the form 
\begin{equation}\label{eq:XHXAdef}
   X_{H,A} = \left( 1 + \varrho_{H,A}\,e^{i\varphi_{H,A}} \right) \ln\frac{m_B}{\Lambda_h} \,,
\end{equation}
with $\Lambda_h=0.5$\,GeV, $0\le\varrho_H, \varrho_A\le 1$, and arbitrary strong-interaction phases $\varphi_H$ and $\varphi_A$. For a detailed exposition of the formalism of QCD factorization and its phenomenology, the reader is referred to \cite{Beneke:2003zv}. 
 
\subsection{SM predictions}
\label{sec:4.1}

Using this phenomenological approach, the CP-averaged branching fractions for the five decay modes entering our ratios are predicted as
\begin{align}
   \text{Br}(B_d\to K^0\bar K^0)
   &= \big( 1.14\,{}_{-0.06}^{+0.06}\,{}_{-0.16}^{+0.22}\,{}_{-0.07}^{+0.03}\,{}_{-0.47}^{+1.18} \big) \cdot 10^{-6}
    = \big( 1.14\,{}_{-0.50}^{+1.20} \big) \cdot 10^{-6} \,, \notag\\
   \text{Br}(B_s\to K^0\bar K^0)
   &= \big( 28.7\,{}_{-1.0}^{+1.0}\,{}_{-5.1}^{+6.5}\,{}_{-1.8}^{+0.8}\,{}_{-13.2}^{+34.8} \big) \cdot 10^{-6}
    = \big( 28.7\,{}_{-14.3}^{+35.5} \big) \cdot 10^{-6} \,, \notag\\
   \text{Br}(B_s\to K^+ K^-)
   &= \big( 25.7\,{}_{-1.0}^{+1.1}\,{}_{-4.4}^{+5.7}\,{}_{-0.8}^{+0.3}\,{}_{-12.0}^{+32.5} \big) \cdot 10^{-6}
    = \big( 25.7\,{}_{-12.8}^{+33.0} \big) \cdot 10^{-6} \,, \\
   \text{Br}(B^+\to\pi^+ K^0)
   &= \big( 17.9\,{}_{-0.6}^{+0.6}\,{}_{-5.9}^{+7.1}\,{}_{-1.1}^{+0.4}\,{}_{-\phantom{1}6.4}^{+15.0} \big) \cdot 10^{-6}
    = \big( 17.9\,{}_{-\phantom{1}8.8}^{+16.6} \big) \cdot 10^{-6} \,, \notag\\
   \text{Br}(B_d\to\pi^- K^+)
   &= \big( 14.7\,{}_{-0.6}^{+0.6}\,{}_{-4.7}^{+5.7}\,{}_{-0.4}^{+0.2}\,{}_{-\phantom{1}5.4}^{+13.1} \big) \cdot 10^{-6}
    = \big( 14.7\,{}_{-\phantom{1}7.2}^{+14.2} \big) \cdot 10^{-6} \,, \notag
\end{align}
and for the direct and the CP asymmetries in the decay $B_s\to K^+ K^-$ one obtains
\begin{equation}
\begin{aligned}
   C_{K^+ K^-}^s
   &= \big(+ 6.98\,{}_{-0.14}^{+0.14}\,{}_{-1.85}^{+1.38}\,{}_{-0.09}^{+0.03}\,{}_{-14.10}^{+12.25} \big)\cdot 10^{-2}
    = +0.07\,{}_{-0.14}^{+0.12} \,, \\
   S_{K^+ K^-}^s
   &= +0.339\,{}_{-0.007}^{+0.007}\,{}_{-0.025}^{+0.019}\,{}_{-0.009}^{+0.004}\,{}_{-0.105}^{+0.117} 
    = +0.34\,{}_{-0.11}^{+0.12} \,.
\end{aligned}
\end{equation}
Here and below, the first error results from the variation of the CKM parameters $|V_{cb}|$, $|V_{ub}|$, and $\gamma$, the second error refers to variations of the renormalization scale, quark masses, meson decay constants, and transition form factors, and the third error corresponds to the uncertainty due to the Gegenbauer moments in the expansion of the LCDAs. The last error reflects an estimate of unknown power corrections, as parameterized by the quantities $X_H$ and $X_A$. In all cases, this last uncertainty dominates the error budget by a large margin. It is important to stress at this point that the theoretical error estimates of unknown power corrections and scale uncertainties should not be treated as Gaussian, but rather as an educated guess of the ranges in which the true values of the observables are expected with high confidence.

Within the sizable theoretical uncertainties, the above predictions are compatible with the experimental values \cite{ParticleDataGroup:2024cfk}
\begin{align}\label{eq:BRsexp}
   \text{Br}(B_d\to K^0\bar K^0)_{\rm exp}
   &= (1.21\pm 0.16)\cdot 10^{-6} \,, \notag\\
   \text{Br}(B_s\to K^0\bar K^0)_{\rm exp}
   &= (17.6\pm 3.1)\cdot 10^{-6} \,, \notag\\
   \text{Br}(B_s\to K^+ K^-)_{\rm exp}
   &= (27.2\pm 2.3)\cdot 10^{-6} \,, \\
   \text{Br}(B^+\to\pi^+ K^0)_{\rm exp}
   &= (23.9\pm 0.6)\cdot 10^{-6} \,, \notag\\
   \text{Br}(B_d\to\pi^- K^+)_{\rm exp}
   &= (20.0\pm 0.4)\cdot 10^{-6} \,, \notag
\end{align}
and 
\begin{equation}\label{eq:CSvals}
\begin{aligned}
   C_{K^+ K^-}^{s,{\rm exp}}
   &= +0.162\pm 0.035 \,, \\
   S_{K^+ K^-}^{s,{\rm exp}}
   &=+ 0.14\pm 0.05 \,.
\end{aligned}
\end{equation}
Considering the central values alone, the largest ``tensions'' concern the branching fractions for the decays $B_s\to K^0\bar K^0$ and $B\to \pi K$ as well as two CP asymmetries, but given the size of the theoretical uncertainties these deviations are not significant.

Many of the dominant theoretical errors cancel in ratios of branching fractions. For the SM predictions for the three ratios in \eqref{eq:defR}, we find
\begin{equation}\label{eq:ratiosQCDF}
\begin{aligned}
   R_{KK}^{sd} 
   &= 1.100\,{}_{-0.003}^{+0.003}\,{}_{-0.146}^{+0.156}\,{}_{-0.075}^{+0.073}\,{}_{-0.091}^{+0.098} 
    = 1.10\,{}_{-0.19}^{+0.20} \,, \\
   R_{KK}^{ss} 
   &= 1.117\,{}_{-0.021}^{+0.021}\,{}_{-0.013}^{+0.007}\,{}_{-0.039}^{+0.016}\,{}_{-0.032}^{+0.028} 
    = 1.12\,{}_{-0.06}^{+0.04} \,, \\
   R_{\pi K}^{ud}
   &= 1.128\,{}_{-0.023}^{+0.023}\,{}_{-0.019}^{+0.012}\,{}_{-0.037}^{+0.015}\,{}_{-0.034}^{+0.029} 
    = 1.13\,{}_{-0.06}^{+0.04} \,.
\end{aligned}
\end{equation}
These values of the three ratios in \eqref{eq:ratiosQCDF} are all significantly larger than the experimental ones given in \eqref{eq:expR}. The remaining uncertainties, as estimated within the QCD factorization framework, are significantly larger for the first ratio than for the other ones. This is in line with the argument presented in \cite{Amhis:2022hpm} that final-state rescattering and other subleading effects can significantly alter the value of $R_{KK}^{sd}$, while they are expected to have a lesser impact on the other two ratios. Nevertheless, within the QCD factorization approach it would be difficult to explain the central experimental value for $R_{KK}^{sd}$ as well. For the two related observables, we obtain 
\begin{equation}
\begin{aligned}
   R_{KK}^{ss} - S_{K^+ K^-}^s \cot\gamma
   &= 1.043\,{}_{-0.003}^{+0.003}\,{}_{-0.011}^{+0.005}\,{}_{-0.039}^{+0.016}\,{}_{-0.170}^{+0.184} 
    = 1.04\,{}_{-0.17}^{+0.19} \,, \\
   R_{KK}^{ss}/R_{\pi K}^{ud}
   &= 0.990\,{}_{-0.001}^{+0.001}\,{}_{-0.009}^{+0.011}\,{}_{- 0.037}^{+0.036}\,{}_{-0.009}^{+0.006} 
    = 0.99\,{}_{-0.04}^{+0.04} \,.
\end{aligned}
\end{equation}
They also differ significantly from the corresponding experimental values shown in \eqref{eq:CSvals}.

\subsection{NP effects in the SM operator basis}
\label{sec:4.2}

We have shown in Section~\ref{sec:bascon} that NP explanations of the three puzzles must involve isospin-violating non-spectator contributions to the decay amplitudes, parameterized by $b_1$, $b_2$ and $b_{3,{\rm EW}}^p$, $b_{4,{\rm EW}}^p$ in \eqref{eq:amplitudes}. Within the operator basis of the SM, this can be done by modifying the Wilson coefficients of the current--current operators $Q_1$ and $Q_2$ (entering $b_1$ and $b_2$) and of the electroweak penguin operators $Q_i$ with $i=7,8,9,10$. Given that the measured value of the ratio $R_{\pi K}^{ud}$ in \eqref{eq:expR} is close to the theoretical prediction obtained using QCD factorization, see \eqref{eq:ratiosQCDF}, a preferred solution would be one that modifies $R_{KK}^{ss}$ directly. Examining \eqref{eq:amplitudes}, there are exactly two non-spectator contributions that could do so, namely \cite{Beneke:2003zv}
\begin{equation}
\begin{aligned}
   b_1 &= \frac{C_F}{N_c^2}\,C_1\spac A_1^i \,, \\
   b_{4,{\rm EW}}^p &= \frac{C_F}{N_c^2} \left( C_{10}\spac A_1^i + C_8\spac A_2^i \right) .
\end{aligned}
\end{equation}
Another, more contrived way to achieve our goal is to invoke NP in operators that affect $R_{\pi K}^{ud}$ directly (in combination with other ones that shift both $R_{\pi K}^{ud}$ and $R_{KK}^{ss}$ in a similar way). Examining \eqref{eq:amplitudes}, the two non-spectator contributions that could do so are
\begin{equation}
\begin{aligned}
   b_2 &= \frac{C_F}{N_c^2}\,C_2\spac A_1^i \,, \\
   b_{3,{\rm EW}} &= \frac{C_F}{N_c^2} \left[ C_9\spac A_1^i 
    + C_7\,\big( A_3^i + A_3^f \big) + N_c\,C_8\spac A_3^f \right] .
\end{aligned}
\end{equation}
Here $N_c=3$, $C_F=4/3$, and the expressions for $A_n^i$ and $A_3^f$ can be found in equation (54) of \cite{Beneke:2003zv}. The superscript $i$ or $f$ refers to gluon emission from an initial- or final-state quark, and the subscript refers to different Dirac structures of the four-fermion operators. The coefficients $C_i$ are the Wilson coefficients of the various operators in the effective weak Hamiltonian \cite{Buchalla:1995vs}. There is a large uncertainty in the calculation of the quantities $A_n^{i,f}$, which involve endpoint-divergent convolution integrals over light-meson LCDAs. For our numerical estimates, we use the approximate expressions 
\begin{equation}\label{eq:Akernels}
\begin{aligned}
   A_1^i &\approx A_2^i \approx 2\pi\alpha_s \left[ 9 \left( X_A - 4 + \frac{\pi^2}{3} \right)
    + r_\chi^{M_1}\spac r_\chi^{M_2}\spac X_A^2 \right] , \\[-1mm]
   A_3^i &\approx 6\pi\alpha_s \left( r_\chi^{M_1} - r_\chi^{M_2} \right)
    \left( X_A^2 - 2 X_A + \frac{\pi^2}{3} \right) , \\[1mm]
   A_3^f &\approx 6\pi\alpha_s \left( r_\chi^{M_1} + r_\chi^{M_2} \right) \left( 2 X_A^2 - X_A \right) ,
\end{aligned}
\end{equation}
derived in (57) of \cite{Beneke:2003zv}, with $X_A$ given in \eqref{eq:XHXAdef}. Depending on the choice of $X_A$ the values of these quantities can be large, up to $\mathcal{O}(100)$ for $A_1^i$ and $A_3^f$, but they multiply the small ratios in \eqref{eq:BKKoverAKK}, so their overall effect is still suppressed compared with the $\alpha_i$ terms in \eqref{eq:amplitudes}.

The Wilson coefficients $C_{1,2}$ of the current--current operators and $C_{7,8,9,10}$ of the electroweak penguin operators also affect other terms in the decay amplitudes \eqref{eq:amplitudes}, as parameterized by the $\alpha_i$ coefficients. The precise dependencies are complicated and involve convolution integrals of perturbative hard-scattering kernels with meson LCDAs. The detailed expressions can be found in Section~2.4 of \cite{Beneke:2003zv}. For our purposes here, it will be sufficient to use the central values for all input parameters. We then obtain
\begin{equation}
\begin{aligned}
   \alpha_1 &\approx \left( C_1 + \rho_2\spac C_2 \right) , \\
   \alpha_4^p &\approx \rho_4^p\,C_1 + \text{terms involving other coefficients} \,, \\
   \alpha_{4,{\rm EW}}^p &\approx \left( C_{10} + \rho_9\spac C_9 \right) + 1.42 \left( C_8 + \rho_7\spac C_7 \right) 
    + \rho_{4,{\rm EW}}^p \left( C_1 + 3\spac C_2 \right) , 
\end{aligned}
\end{equation}
where for $K\bar K$ and $\pi\bar K$ final states
\begin{equation}
\begin{aligned}
   \rho_2^{K\bar K} &= \rho_9^{K\bar K} = 0.34 - 0.08\spac i \,, && \\
   \rho_2^{\pi\bar K} &= \rho_9^{\pi\bar K} = 0.32 - 0.08\spac i \,, && \\
   \rho_7 &= 0.29 \,, \\
   \rho_4^c &= - 0.03 - 0.03\spac i \,, & \rho_4^u &= -0.02 - 0.04\spac i \,, \\
   \rho_{4,{\rm EW}}^c &= -0.0003 - 0.0003\spac i \,, \quad & \rho_{4,{\rm EW}}^u &= -0.0002 - 0.0005\spac i \,.
\end{aligned}
\end{equation}
Given the smallness of the ratios in \eqref{eq:BKKoverAKK}, the effect of $\alpha_1$ on maintaining the equality $R_{KK}^{ss}=R_{\pi K}^{ud}$ is, in general, much stronger than the effect of $b_1$ or $b_2$ on violating it, and the same is true for $\alpha_{4,{\rm EW}}$ vs.\ $b_{3,{\rm EW}}$ or $b_{4,{\rm EW}}$. These observations lead to the following conclusions:
\begin{itemize}
\item 
If one invokes NP contributions to one or more of the electroweak penguin coefficients to account for the experimental data, their values must be fine-tuned such that the combination $|(C_{10}^{\rm NP}+\rho_9\spac C_9^{\rm NP})+1.42\spac(C_8^{\rm NP}+\rho_7\spac C_7^{\rm NP})|$ is much smaller than the power-suppressed weak annihilation contributions proportional to $b_{3,{\rm EW}}$ and $b_{4,{\rm EW}}$.
\item 
If one invokes NP contributions to $C_1$ and $C_2$ to account for the experimental data, their values must be fine-tuned such that $|C_1^{\rm NP}+\rho_2\,C_2^{\rm NP}|$ is much smaller than the power-suppressed weak annihilation contributions proportional to $b_1$ and $b_2$.
\end{itemize}

\begin{figure}
\begin{center}
\includegraphics[width=0.45\textwidth]{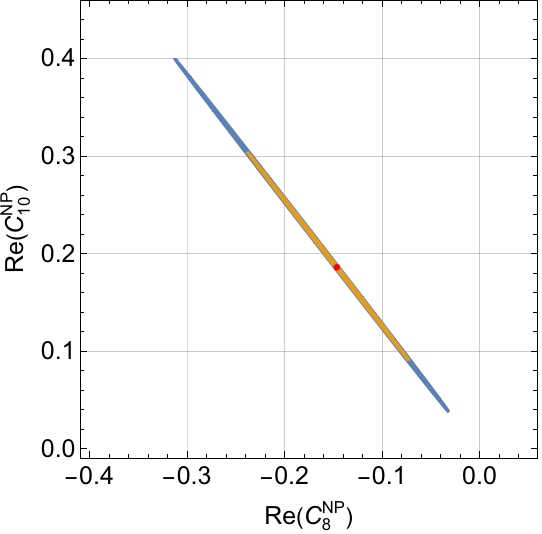} \quad 
\includegraphics[width=0.456\textwidth]{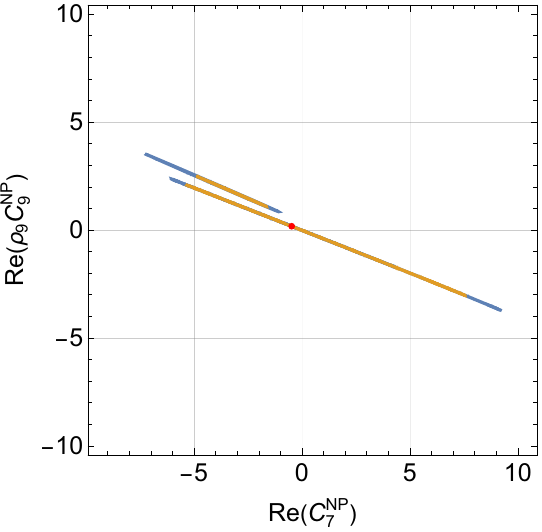}
\end{center}
\vspace{-3mm}
\caption{The $1\sigma$ (orange) and $2\sigma$ (blue) allowed regions for ${\rm Re}(C_8^{\rm NP})$ vs.\ ${\rm Re}(C_{10}^{\rm NP})$ (left panel) and ${\rm Re}(C_7^{\rm NP})$ vs.\ ${\rm Re}(\rho_9\spac C_9^{\rm NP})$ (right panel) that can account for the experimental values of $R_{KK}^{ss}$ and $R_{\pi K}^{ud}$, with the imaginary parts held fixed at the best fit values. All values refer to the scale $\mu=m_b$. The red dot represents the best fit.}
\label{fig:SMbasisfit1}
\end{figure}

\begin{figure}
\begin{center}
\includegraphics[width=0.45\textwidth]{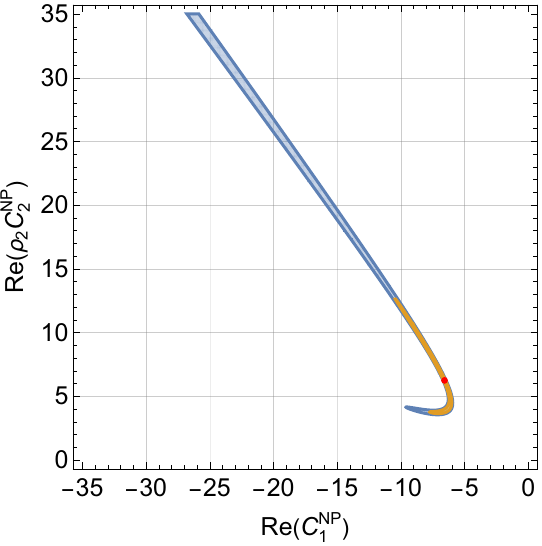}
\end{center}
\vspace{-3mm}
\caption{The $1\sigma$ (orange) and $2\sigma$ (blue) allowed regions for ${\rm Re}(C_1^{\rm NP})$ vs.\ ${\rm Re}(\rho_2\spac C_2^{\rm NP})$ that can account for the experimental values of $R_{KK}^{ss}$ and $R_{\pi K}^{ud}$, with the imaginary parts held fixed at the best fit values. All values refer to the scale $\mu=m_b$. The red dot represents the best fit.}
\label{fig:SMbasisfit2}
\end{figure}

We have performed three fits to the ratios $R_{KK}^{ss}$ and $R_{\pi K}^{ud}$, each time considering a pair of operators as the only NP contributions, namely $(C_8^{\rm NP}, C_{10}^{\rm NP})$, $(C_7^{\rm NP}, C_9^{\rm NP})$, and $(C_1^{\rm NP}, C_2^{\rm NP})$. Other combinations of coefficients would yield similar results. Our findings are summarized in Figures~\ref{fig:SMbasisfit1} and \ref{fig:SMbasisfit2}. We obtain the central values
\begin{equation}
\begin{aligned}
   &\text{Fit 1:} \quad & C_8^{\rm NP} &= - 0.15 - 0.018\spac i \,, \quad & C_{10}^{\rm NP} &= 0.19 + 0.030\spac i \,, \\
   &\text{Fit 2:} & C_7^{\rm NP} &= - 0.47 - 0.83\spac i \,, & C_{9}^{\rm NP} &= 0.21 + 1.2\spac i \,, \\
   &\text{Fit 3:} & C_1^{\rm NP} &= - 6.5 + 5.3\spac i \,, & C_2^{\rm NP} &= 23 - 17\spac i \,.
\end{aligned}
\end{equation}
In comparison, the SM values of these Wilson coefficients are much smaller, namely
\begin{equation}\label{eq:fitresults}
\begin{aligned}
   C_8^{\rm SM} &\approx 4.7\cdot 10^{-4} \,, & C_{10}^{\rm SM} &\approx 1.7\cdot 10^{-3} \,, \\
   C_7^{\rm SM} &\approx - 8.6\cdot 10^{-5} \,, \quad & C_9^{\rm SM} &\approx -9.7\cdot 10^{-3} \,, \\
   C_1^{\rm SM} &\approx 1.08 \,, & C_2^{\rm SM} &\approx - 0.19 \,.
\end{aligned}
\end{equation}
Associating energy scales with the NP contributions via 
\begin{equation}
\begin{aligned}
   \frac{1}{\left( \Lambda_i^{\rm NP} \right)^2} &\equiv \frac{G_F}{\sqrt2}\,|V_{cb} V_{cs}^*|\,|C_i^{\rm NP}| \,; \quad i=7,8,9,10 \,, \\
   \frac{1}{\left( \Lambda_i^{\rm NP} \right)^2} &\equiv \frac{G_F}{\sqrt2}\,|V_{ub} V_{us}^*|\,|C_i^{\rm NP}| \,; \quad i=1,2 \,,
\end{aligned}
\end{equation}
the central values in \eqref{eq:fitresults} correspond to
\begin{equation}
\begin{aligned}
   &\text{Fit 1:} \quad & \Lambda_8^{\rm NP} &\approx 4.5\,\text{TeV} \,, \quad & \Lambda_{10}^{\rm NP} &\approx 4.0\,\text{TeV} \,, \\
   &\text{Fit 2:} & \Lambda_7^{\rm NP} &\approx 1.8\,\text{TeV} \,, \quad & \Lambda_9^{\rm NP} &\approx 1.5\,\text{TeV} \,, \\
    &\text{Fit 3:} &\Lambda_1^{\rm NP} &\approx 4.2\,\text{TeV} \,, \quad & \Lambda_2^{\rm NP} &\approx 2.3\,\text{TeV} \,.
\end{aligned}
\end{equation}
These scales are much lower than the scale of 5.6\,TeV characteristic for $|\Delta B|=|\Delta S|=1$ transitions in the SM, see \eqref{eq:SMscale}, indicating that these solutions are not only fine-tuned but also highly unnatural.

Two points, which make the solutions invoking NP contributions to the electroweak penguin operators rather implausible, are worth emphasizing:
\begin{itemize}
\item 
The leading-order correction to $R_{\pi K}^{ud}$ is proportional to a linear combination of the coefficients $\alpha_{4,{\rm EW}}^c$ and $b_{3,{\rm EW}}^c$. The fine-tuning apparent in Figure~\ref{fig:SMbasisfit1} corresponds to
\begin{align}
   \left| \text{Re}\!\left( \alpha_{4,{\rm EW}}^c - \frac{B_{\pi\bar K}}{A_{\pi\bar K}}\,b_{3,{\rm EW}}^c \right)_{\rm NP} \right| 
   &\approx \left| \text{Re}\!\left[ C_{10}^{\rm NP} + \left( 1.42 - \frac{C_F}{N_c}\,A_3^f\,\frac{B_{\pi\bar K}}{A_{\pi\bar K}} \right) 
    C_8^{\rm NP} \right] \right| \notag\\
   &\approx 0.0023 \ll \text{Re}(C_{10}^{\rm NP}) \approx 0.19
\end{align}
for the fit in the $(C_8^{\rm NP}, C_{10}^{\rm NP})$ plane, where the numbers in the last line refer to the best fit values. A similar relation, indicating another fine tuning by two orders of magnitude, can be derived for the fit in the $(C_7^{\rm NP}, C_9^{\rm NP})$ plane. This confirms our expectation that the shift of $R_{\pi K}^{ud}$ due to $\alpha_{4,{\rm EW}}^c$ and $b_{3,{\rm EW}}^c$ must be fine-tuned to be small. 
\item 
The proximity of $R_{\pi K}^{ud}$ to~1, rather than being a natural consequence of $\alpha_4^c$ dominance, is now an accidental result, because the shift by $\frac32\spac(\alpha_{4,{\rm EW}}^c/\alpha_4^c)_{\rm NP}\approx 0.3$, which brings $R_{KK}^{ss}$ close to its experimental value, affects $R_{\pi K}^{ud}$ in the same way, but cancels in the latter via non-spectator contributions.
\end{itemize}

For the fit in the $(C_1^{\rm NP}, C_2^{\rm NP})$ plane, the fact that it yields huge NP contributions to the topological tree amplitudes is a reason for concern. Constraints on NP contributions to $C_1$ and $C_2$ from the width splitting in the $B_s$ system, $\Delta\Gamma_s$, were studied in \cite{Lenz:2019lvd}. The bounds of $\mathcal{O}(1)$ are based, however, on the contribution to the $b\to c\bar cs$ transition and thus do not apply to our scenario. The bounds on the contribution to the $b\to u\bar us$ transition can be estimated to be a factor of ${\cal O}(100)$ weaker and thus do not disfavor the $(C_1^{\rm NP}, C_2^{\rm NP})$ solution per se. However, the CP-averaged rates for the decays $B^-\to\pi^0 K^-$ and $\bar B_d\to\pi^0\bar K^0$ involving neutral pions are sensitive to the orthogonal combination $\alpha_2\approx C_2+\rho_2\spac C_1$, and they can be used to firmly exclude the $(C_1^{\rm NP}, C_2^{\rm NP})$ scenario. Indeed, varying the parameters in the QCD factorization approach as described earlier, we find for this scenario
\begin{equation}
\begin{aligned}
   \text{Br}(B^+\to\pi^0 K^+)
   &= \big( 438\,{}_{-63}^{+97} \big) \cdot 10^{-6} \,, \\
   \text{Br}(B_d\to\pi^0 K^0)
   &= \big( 510\,{}_{-55}^{+60} \big) \cdot 10^{-6} \,,
\end{aligned}
\end{equation}
where for simplicity we have combined all theoretical uncertainties into a single error. These rates are almost two orders of magnitude larger than the experimental values \cite{ParticleDataGroup:2024cfk}
\begin{equation}\label{eq:Kpi0rates}
\begin{aligned}
   \text{Br}(B^+\to\pi^0 K^+)_{\rm exp}
   &= \big( 13.2\pm 0.4 \big) \cdot 10^{-6} \,, \\
   \text{Br}(B_d\to\pi^0 K^0)_{\rm exp}
   &= \big( 10.1\pm 0.4 \big) \cdot 10^{-6} \,.
\end{aligned}
\end{equation}

\subsection{Flavor-specific NP contributions}

We now turn to more general extensions of the SM featuring flavor-specific NP operators. We have calculated the explicit expressions for the flavor-specific NP contributions in \eqref{eq:TNP} at NLO in the QCD factorization approach, following the procedure outlined in \cite{Beneke:2003zv} and applying it to the case where all mesons are pseudoscalars. The NP contributions to the {\bf tree topologies} are contained in the parameters
\begin{align}\label{eq:treeeffects}
   \alpha_{1,q}^{D,\spac{\rm NP}}(M_1 M_2)
   &= C_{4,q}^D + \frac{C_{3,q}^D}{N_c} \left\{ 1
    + \frac{C_F\spac\alpha_s}{4\pi} \left[ V_4(M_2) + \frac{4\pi^2}{N_c}\,H_4(M_1 M_2) \right] \right\} \notag\\
   &\quad + r_\chi^{M_2} \left[ C_{6,q}^D + \frac{C_{5,q}^D}{N_c} \left\{ 1
    + \frac{C_F\spac\alpha_s}{4\pi} \left[ V_6(M_2) + \frac{4\pi^2}{N_c}\,H_6(M_1 M_2) \right] \right\} \right] , \notag\\
   \alpha_{2,q}^{D,\spac{\rm NP}}(M_1 M_2)
   &= C_{3,q}^D + \frac{C_{4,q}^D}{N_c} \left\{ 1
    + \frac{C_F\spac\alpha_s}{4\pi} \left[ V_3(M_2) + \frac{4\pi^2}{N_c}\,H_3(M_1 M_2) \right] \right\} \notag\\
   &\quad - \left[ C_{5,q}^D + \frac{C_{6,q}^D}{N_c} \left\{ 1
    + \frac{C_F\spac\alpha_s}{4\pi} \left[ V_5(M_2) + \frac{4\pi^2}{N_c}\,H_5(M_1 M_2) \right] \right\} \right] .
\end{align}
The ratios $r_\chi^\pi$ and $r_\chi^K$ have been defined in \eqref{eq:rchidef}. The explicit expressions for the vertex corrections $V_i(M_2)$ and the hard spectator interactions $H_i(M_1 M_2)$ are given in the Appendix. Some of the convolution integrals involving chirally-enhanced power corrections to the kernels $H_i(M_1 M_2)$ contain endpoint divergences, which we isolate and express in terms of the parameters $X_H$ in \eqref{eq:XHXAdef}. The NP contributions to the {\bf penguin topology} are described by the parameters
\begin{equation}\label{eq:penguineffects}
   \alpha_4^{D,\spac{\rm NP}}(M_1 M_2)
   =  \sum_{q=u,d,s,c,b} \big[ P_{4,q}(M_2) + r_\chi^{M_2}\spac P_{6,q}(M_2) \big] \,,
\end{equation}
whose explicit form is also given in the Appendix. Finally, the NP contributions to the {\bf annihilation topologies} are accounted for by the parameters
\begin{equation}\label{eq:annihilationeffects}
\begin{aligned}
   b_{1,q}^{\rm NP}(M_1 M_2)
   &= \frac{C_F}{N_c^2} \left( C_{4,q}^D\spac A_1^i + C_{6,q}^D\spac A_2^i \right) , \\
   b_{2,q}^{\rm NP}(M_1 M_2)
   &= \frac{C_F}{N_c^2} \left[ C_{3,q}^D\spac A_1^i + C_{5,q}^D\,\big( A_3^i + A_3^f \big)
    + N_c\,C_{6,q}^D\spac A_3^f \right] ,
\end{aligned}
\end{equation}
where we use the approximate expressions \eqref{eq:Akernels} for $A_n^i$ and $A_3^f$.

Most scale-dependent quantities in \eqref{eq:treeeffects}, \eqref{eq:penguineffects}, and \eqref{eq:annihilationeffects} are evaluated at a scale $\mu_b$ of $\mathcal{O}(m_b)$. However, for the hard spectator scattering and weak annihilation terms, we set $\mu$ to an intermediate ``hard-collinear'' scale $\mu_{hc}$ of $\mathcal{O}(\sqrt{m_b\spac\Lambda_{\rm QCD}})$. In assessing the scale uncertainties of our results, we vary the scales by a factor~2 about their default values.

\subsection{Numerical results for the flavor-specific NP effects}

Assuming that the NP contributions are subdominant compared with the SM amplitudes, it makes sense to linearize the results for the three ratios in the NP Wilson coefficients. Keeping the terms with the eight largest coefficients in magnitude (for both $|\Delta S|=1$ $|\Delta D|=1$ transitions), we find 
\begin{equation}
\begin{aligned}
   R_{KK}^{sd}
   &\simeq 1.13 - 0.80\,\text{Re}\big(C_{6,d}^s\big)
    - 0.56\,\text{Re}\big(C_{4,d}^s\big)
    - 0.23\,\text{Re}\big(C_{5,d}^s\big)
    - 0.19\,\text{Re}\big(C_{3,d}^s\big) \\  
   &\quad - 0.08\,\text{Re}\big(C_{6,s}^s\big)
    + 0.03\,\text{Re}\big(C_{4,c}^s\big)
    + 0.03\,\text{Re}\big(C_{6,c}^s\big)
    + 0.02\,\text{Re}\big(C_{4,u}^s\big) 
    + \dots \\
   &\quad - 3.53\,\text{Re}\big(C_{6,s}^d\big)
    - 2.48\,\text{Re}\big(C_{4,s}^d\big)
    - 1.34\,\text{Im}\big(C_{6,s}^d\big)
    - 1.03\,\text{Re}\big(C_{5,s}^d\big) \\
   &\quad -0.94\,\text{Im}\big(C_{4,s}^d\big) 
    - 0.78\,\text{Re}\big(C_{3,s}^d\big)
    - 0.39\,\text{Im}\big(C_{5,s}^d\big)
    - 0.29\,\text{Im}\big(C_{3,s}^d\big)
    + \dots \,,
\end{aligned}
\end{equation}
\begin{equation}\label{Rss linear approx}
\begin{aligned}
   R_{KK}^{ss}
   &\simeq 1.12 + 0.85\,\text{Re}\big(C_{6,u}^s\big) 
    - 0.82\,\text{Re}\big(C_{6,d}^s\big)
    + 0.61\,\text{Re}\big(C_{4,u}^s\big)
    - 0.58\,\text{Re}\big(C_{4,d}^s\big) \\
   &\quad + 0.24\,\text{Re}\big(C_{5,u}^s\big)
    - 0.23\,\text{Re}\big(C_{5,d}^s\big)
    + 0.19\,\text{Re}\big(C_{3,u}^s\big)
    - 0.19\,\text{Re}\big(C_{3,d}^s\big)
    + \dots \,,
\end{aligned}
\end{equation}
and
\begin{equation}\label{Rud linear approx}
\begin{aligned}
   R_{\pi K}^{ud}
   &\simeq 1.13 + 0.82\,\text{Re}\big(C_{6,u}^s\big)
    - 0.78\,\text{Re}\big(C_{6,d}^s\big)
    + 0.64\,\text{Re}\big(C_{4,u}^s\big)
    - 0.61\,\text{Re}\big(C_{4,d}^s\big) \\
   &\quad + 0.23\,\text{Re}\big(C_{5,u}^s\big) 
    - 0.22\,\text{Re}\big(C_{5,d}^s\big)
    + 0.18\,\text{Re}\big(C_{3,u}^s\big)
    - 0.18\,\text{Re}\big(C_{3,d}^s\big)
    + \dots \,, 
\end{aligned}
\end{equation}
where as stated earlier we use $\Lambda_{\rm NP}=10$\,TeV in the normalization of the NP contributions. The Wilson coefficients are normalized at $\mu=m_b$, and for simplicity we neglect their scale variation when computing the decay rates. Note that it would not be difficult to change the ratio $R_{KK}^{sd}$ by NP in such a way that it takes on a value close to 0.62, while keeping the other two ratios largely unaffected, e.g.\ by choosing $C_{6,u}^s\approx C_{6,d}^s\approx 0.63$, or by switching on some $|\Delta D|=1$ Wilson coefficients, e.g.\ $C_{6,s}^d\approx 0.14$. On the contrary, even with generic flavor-specific NP contributions, it seems extremely challenging to modify the ratio $R_{KK}^{ss}$ while keeping $R_{\pi K}^{ud}$ largely unaffected, as the NP contributions in both ratios involve almost identical coefficients. Indeed, for the double ratio we obtain
\begin{align}\label{Rss/Rud linear approx}
   \frac{R_{KK}^{ss}}{R_{\pi K}^{ud}}
   &\simeq 0.990 - 0.040\,\text{Re}\big(C_{6,d}^s\big)
    + 0.033\,\text{Re}\big(C_{6,u}^s\big)
    - 0.024\,\text{Re}\big(C_{4,u}^s\big) 
    + 0.022\,\text{Re}\big(C_{4,d}^s\big) \notag\\[-2.5mm] 
   &\quad + 0.017\,\text{Im}\big(C_{6,s}^s\big) 
    - 0.016\,\text{Im}\big(C_{6,u}^s\big) 
    - 0.014\,\text{Im}\big(C_{6,d}^s\big) 
    - 0.012\,\text{Re}\big(C_{5,d}^s\big) + \dots \,.
\end{align}
This result involves much smaller coefficients in front of the NP terms than the individual ratios. We conclude that, with modest NP coefficients less than $\mathcal{O}(1)$ in magnitude, it is impossible to account for the experimental value shown in \eqref{eq:expsd-ud}.

\begin{figure}
\begin{center}
\includegraphics[scale=0.82]{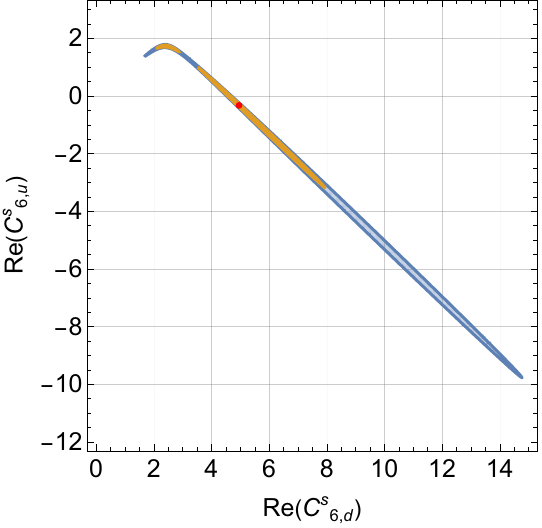} \quad 
\includegraphics[scale=0.8]{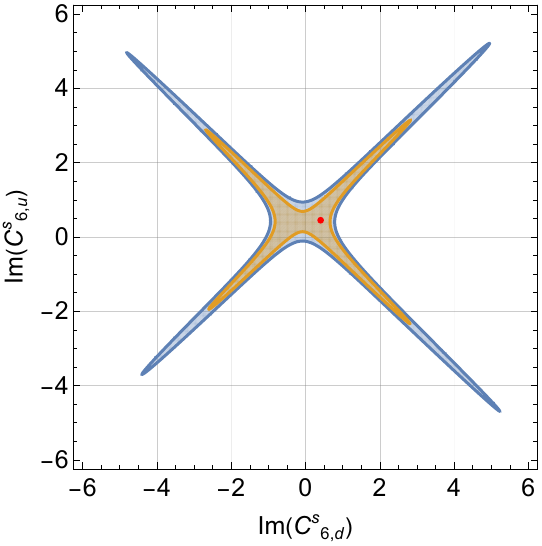} 
\end{center}
\vspace{-3mm}
\caption{The $1\sigma$ (orange) and $2\sigma$ (light blue) allowed regions for ${\rm Re}(C_{6,u}^s)$ vs.\ ${\rm Re}(C_{6,d}^s)$ (left panel) and ${\rm Im}(C_{6,u}^s)$ vs.\ ${\rm Im}(C_{6,d}^s)$ (right panel) that can account for the experimental values of $R_{KK}^{ss}$ and $R_{\pi K}^{ud}$, with the imaginary part (left) and real parts (right) held fixed at the best fit values. All values refer to the scale $\mu=m_b$. The red dot represents the best fit.} 
\label{fig:fit_to_ratios}
\end{figure}

\begin{figure}
\begin{center}
\includegraphics[scale=0.8]{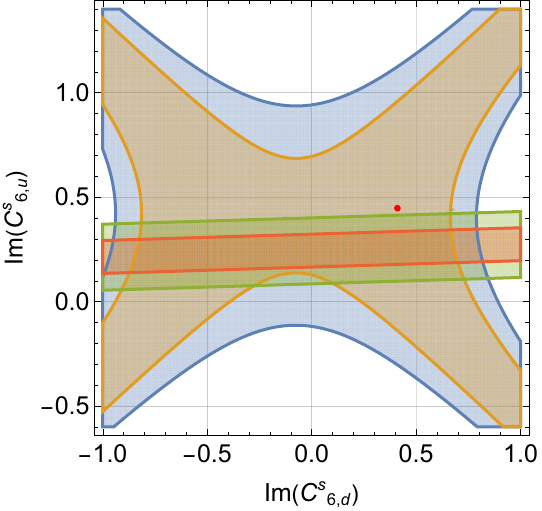} 
\end{center}
\vspace{-3mm}
\caption{The $1\sigma$ (orange) and $2\sigma$ (light blue) allowed regions for ${\rm Im}(C_{6,u}^s)$ vs.\ ${\rm Im}(C_{6,d}^s)$ that can account for the experimental values of $R_{KK}^{ss}$ and $R_{\pi K}^{ud}$, with the real parts held fixed at the best fit values. The overlayed bands show the $1\sigma$ (red) and $2\sigma$ (green) experimental ranges for the mixing-induced CP asymmetry $S_{K^+ K^-}^s$ in $B_s\to K^+ K^-$ decay.} 
\label{fig:SKK_from_fit}
\end{figure}

The situation changes when the exact theoretical expressions are used for the decay rates, which are non-linear in the NP coefficients. In the following discussion, we will concentrate on the two ratios $R_{KK}^{ss}$ and $R_{\pi K}^{ud}$, since they are purely sensitive to NP effects in $|\Delta B|=|\Delta S|=1$ transitions. We have performed a four-parameter fit to these observables in terms of the real and imaginary parts of the two NP coefficients $C_{6,u}^s$ and $C_{6,d}^s$ (again at the scale $\mu=m_b$), which enter with the largest coefficients in the above expressions. (Analogous conclusions could be drawn by picking another pair of Wilson coefficients.) In Figure~\ref{fig:fit_to_ratios}, we show the result of the fit in the plane of the real parts (left panel) and imaginary parts (right panel) of $C_{6,u}^s$ and $C_{6,d}^s$, keeping the imaginary parts (left) and real parts (right) fixed at the best fit values
\begin{equation}\label{eq:centralfitvalues}
   C_{6,u}^s = - 0.33 + 0.45\spac i \,, \qquad
   C_{6,d}^s = 4.96 + 0.41\spac i \,.
\end{equation}
The allowed regions at $1\sigma$ and $2\sigma$ are shown together with the best fit point. The solution for the real parts of the two coefficients is again fine-tuned. The reason can be inferred from the structure of the NP contributions to the relevant decay amplitudes in \eqref{eq:NPamplitudes}. The ratio $R_{KK}^{ss}$ can be made smaller than unity by means of different NP parameters $\alpha_{1,u}^{s,\spac{\rm NP}}\ne\alpha_{1,d}^{s,\spac{\rm NP}}$, which according to \eqref{eq:treeeffects} can be done by choosing $C_{6,u}^s\ne C_{6,d}^s$. However, in order to maintain the SM relation $R_{\pi K}\approx 1$, which agrees with the data, one must cancel the contributions of the Wilson coefficients $C_{6,u}^s$ and $C_{6,d}^s$ to $(\alpha_{1,u}^{s,\spac{\rm NP}}+\alpha_4^{s,\spac{\rm NP}})$ and $(\alpha_{1,d}^{s,\spac{\rm NP}}+\alpha_4^{s,\spac{\rm NP}})$ in the last two lines of \eqref{eq:NPamplitudes} against the contributions of these coefficients to the power-suppressed weak annihilation terms $b_{2,d}^{s,\spac{\rm NP}}$ and $b_{2,u}^{s,\spac{\rm NP}}$, respectively. Note that the imaginary parts of the Wilson coefficients are less constrained as long as we restrict them to values of $\mathcal{O}(1)$ or less, see Figure~\ref{fig:SKK_from_fit}. The reason is that the imaginary parts enter the expressions for $R_{KK}^{ss}$ and $R_{\pi K}^{ud}$ with smaller prefactors than the real parts.

We have performed the fit to the ratios $R_{KK}^{ss}$ and $R_{\pi K}^{ud}$ using the central values of the predictions for these quantities in the QCD factorization approach. We now present the predictions obtained for the three ratios in \eqref{eq:defR} including an estimate of the theoretical uncertainties as described earlier in Section~\ref{sec:4.1}. We find 
\begin{equation}\label{eq:fitratios}
\begin{aligned}
   R_{KK}^{sd} &= 0.87\,{}_{-0.06}^{+0.06}\,{}_{-0.29}^{+0.27}\,{}_{-0.03}^{+0.08}\,{}_{-0.75}^{+1.29} 
    = 0.87\,{}_{-0.80}^{+1.33} \,, \\
   R_{KK}^{ss} &= 0.65\,_{-0.04}^{+0.05}\,_{-0.18}^{+0.15}\,_{-0.02}^{+0.06}\,_{-0.56}^{+1.00} 
    = 0.65\,_{-0.59}^{+1.01} \,, \\ 
   R_{\pi K}^{ud} &= 1.11\,_{-0.08}^{+0.08}\,_{-0.23}^{+0.20}\,_{-0.04}^{+0.10}\,_{-0.41}^{+0.54}
    = 1.11\,_{-0.48}^{+0.59} \,.
\end{aligned}
\end{equation}
As explained earlier, the first error results from the variation of the CKM parameters, the second error refers to variations of the renormalization scale, quark masses, decay constants, and form factors, the third error corresponds to the variations of the Gegenbauer moments of the LCDAs, and the last error reflects an estimate of unknown power corrections.  The central values of $R_{KK}^{ss}$ and $R_{\pi K}^{ud}$ are in perfect agreement with experiment, see \eqref{eq:expR}. The ratio $R_{KK}^{sd}$, which was not included in the fit, agrees with experiment within the theoretical uncertainties. Its central value could be brought closer to the experimental value by adding some small NP contributions to the $B_d\to K^0\bar K^0$ decay amplitude. The estimated theoretical uncertainties are significantly larger than those in the SM predictions in \eqref{eq:ratiosQCDF}, because some of the terms involving the coefficients $C_{6,u}^s$ and $C_{6,d}^s$ multiply the weak annihilation amplitudes $A_2^i$ and $A_3^f$ in \eqref{eq:annihilationeffects}, which are afflicted by very large uncertainties. 

The fact that $\text{Re}(C_{6,d}^s)\gg 1$ indicates that for the solution considered here some NP effects are larger than the SM contributions to the decay amplitudes. In fact, the effective NP scale $\Lambda_{\rm NP}/|C_{6,d}^s|^{1/2}\approx 4.5\,\text{TeV}$ is below the characteristic scale in \eqref{eq:SMscale} associated with the SM penguin contributions. This raises the question if, even though the ratios in \eqref{eq:fitratios} agree with experiment, the individual CP-averaged branching ratios entering the three ratios are perhaps far off the experimental results shown in \eqref{eq:BRsexp}. Remarkably, we find that this is not the case. The values for the branching fractions we obtain with the central values for the NP coefficients in \eqref{eq:centralfitvalues} are
\begin{equation}
\begin{aligned}
   \text{Br}(B_d\to K^0\bar K^0)
   &= \big( 1.14\,{}_{-0.06}^{+0.06}\,{}_{-0.16}^{+0.22}\,{}_{-0.07}^{+0.03}\,{}_{-0.47}^{+1.18} \big) \cdot 10^{-6} 
    = \big( 1.14\,{}_{-0.50}^{+1.18} \big) \cdot 10^{-6} \,, \\
   \text{Br}(B_s\to K^0\bar K^0)
   &= \big( 22.7\,{}_{-0.8}^{+0.8}\,{}_{-6.1}^{+6.2}\,{}_{-0.6}^{+1.6}\,{}_{-16.4}^{+15.2} \big) \cdot 10^{-6} 
    = \big( 22.7\,{}_{-17.5}^{+16.5} \big) \cdot 10^{-6} \,, \\
   \text{Br}(B_s\to K^+ K^-)
   &= \big( 35.0\,{}_{-1.2}^{+1.2}\,{}_{-5.9}^{+7.4}\,{}_{-0.9}^{+0.4}\,{}_{-14.8}^{+37.6} \big) \cdot 10^{-6}
    = \big( 35.0\,{}_{-16.0}^{+38.3} \big) \cdot 10^{-6} \,, \\
   \text{Br}(B^+\to\pi^+ K^0)
   &= \big( 17.1\,{}_{-0.5}^{+0.5}\,{}_{-6.7}^{+7.9}\,{}_{-0.4}^{+1.1}\,{}_{-10.4}^{+\phantom{1}9.1} \big) \cdot 10^{-6}
    = \big( 17.1\,{}_{-12.4}^{+12.1} \big) \cdot 10^{-6}  \,, \\
   \text{Br}(B_d\to\pi^- K^+)
   &= \big( 14.3\,{}_{-0.6}^{+0.6}\,{}_{-5.3}^{+6.6}\,{}_{-0.4}^{+0.2}\,{}_{-7.6}^{+5.9} \big) \cdot 10^{-6}
    = \big( 14.3\,{}_{-9.3}^{+8.9} \big) \cdot 10^{-6} \,.
\end{aligned}
\end{equation}
They are consistent with the experimental values in \eqref{eq:BRsexp} within errors. As another cross check of our fit solution, we calculate the $B\to\pi K$ decays involving neutral pions in the final state. For their CP-averaged branching fractions, we obtain
\begin{equation}
\begin{aligned}
   \text{Br}(B^+\to\pi^0 K^+)
   &= \big( 6.6\,{}_{-0.3}^{+0.3}\,{}_{-2.7}^{+3.4}\,{}_{-0.7}^{+2.0}\,{}_{-3.0}^{+7.6} \big) \cdot 10^{-6}
    = \big( 6.6\,{}_{-4.1}^{+8.6} \big) \cdot 10^{-6} \,, \\
   \text{Br}(B_d\to\pi^0 K^0)
   &= \big( 7.8\,{}_{-0.2}^{+0.2}\,{}_{-3.1}^{+3.7}\,{}_{-1.0}^{+2.7}\,{}_{-2.9}^{+5.7} \big) \cdot 10^{-6} 
    = \big( 7.8\,{}_{-4.4}^{+7.4} \big) \cdot 10^{-6} \,.
\end{aligned}
\end{equation}
Within errors, these results are also consistent with the experimental values shown in \eqref{eq:Kpi0rates}. 

We finally evaluate the mixing-induced CP asymmetry in $B_s\to K^+ K^-$ decay for the best fit solution in \eqref{eq:centralfitvalues}, finding
\begin{equation}\label{eq:bad_prediction}
   S_{K^+ K^-}^s 
   =+ 0.019\,{}_{-0.007}^{+0.006}\,{}_{-0.020}^{+0.019}\,{}_{-0.012}^{+0.005}\,{}_{-0.008}^{+0.009}
   = +0.019\,{}_{-0.025}^{+0.022} \,.
\end{equation}
This values appears to be in conflict with the experimental value in \eqref{eq:CSvals}. However, this can be fixed e.g.\ by means of a minor shift of the imaginary of $C_{6,u}^s$ from 0.45 to 0.26, which is well within the $1\sigma$ region of the fit shown in Figure~\ref{fig:SKK_from_fit}. This leads to 
\begin{equation}\label{eq:better_predictions}
   S_{K^+ K^-}^s 
   = +0.140\,{}_{-0.006}^{+0.005}\,{}_{-0.018}^{+0.017}\,{}_{-0.011}^{+0.004}\,{}_{-0.037}^{+0.039} 
   = +0.14\,{}_{-0.04}^{+0.04} \,,
\end{equation}
in perfect agreement with the experimental result. In general, the CP asymmetries exhibit a strong dependence on the new weak phases introduced by the NP Wilson coefficients as well as on the poorly known strong-interaction phases of the weak annihilation contributions, which in the NP scenario considered here are significantly enhanced. Since we did not include $S_{K^+ K^-}^s$ in the fit, it is not too surprising that for the central values the asymmetry is not well reproduced. The solution is still consistent with the data, because $S_{K^+ K^-}^s$ can be reproduced for a value inside the $1\sigma$ region of the fit, see Figure~\ref{fig:SKK_from_fit}.

\section{Conclusions}
\label{sec:con}

The measured value of the branching fraction for the rare decay $B_s\to K^0\bar K^0$ is about 40\% lower than the predicted value in the Standard Model, obtained using the approximate $SU(3)$ flavor symmetry of QCD and arguments based on the heavy-quark expansion. If this $3\sigma$ effect is not due to a fluctuation in the data, it calls for the presence of new physics (NP). In this work, we have argued that possible explanations invoking high-scale NP are fine-tuned and rather contrived. The key ingredient in our argument is the approximate equality $R_{KK}^{ss}=R_{\pi K}^{ud}$ of two ratios defined in \eqref{eq:defR}. The basic qualitative argument why NP explanations of the violation of this are fine-tuned is the following: The quark transitions in the $B_s\to K^0\bar K^0$ and $B^+\to\pi^+K^0$ decays are the same: $\bar b\to\bar s d\bar d$. Similarly, the quark transitions in the $B_s\to K^+ K^-$ and $B_d\to\pi^- K^+$ decays are the same: $\bar b\to\bar s u\bar u$. Thus, any violation of the equality between the two ratios of rates, $R_{KK}^{ss}$ and $R_{\pi K}^{ud}$, necessarily involves isospin-breaking non-spectator contributions. However, any operator that contributes differently to $R_{KK}^{ss}$ and $R_{\pi K}^{ud}$ also gives much larger spectator contributions, which act to restore the equality. These latter contributions can only be suppressed by means of fine-tuning.

Another challenge to NP scenarios relates to the relation between the deviation of $R_{KK}^{ss}$ from unity and the CP asymmetry in the decay $B_s\to K^+K^-$. The leading penguin contributions to $B_s\to K^0\bar K^0$ and to $B_s\to K^+ K^-$ are equal in the isospin limit. Thus, a deviation of $R_{KK}^{ss}$ from unity can only come from tree amplitudes contributing to $B_s\to K^+ K^-$ but not to $B_s\to K^0\bar K^0$. However, the size and phase of the latter (tree amplitude) relative to the former (penguin amplitude) is measured by the CP asymmetry in $B_s\to K^+ K^-$, which is well accounted for within the SM. Affecting $R_{KK}^{ss}$ without modifying $S_{K^+ K^-}^s$ by too much is quite hard to achieve.

These qualitative arguments are indeed quantitatively realized by the scenarios we have considered:
\begin{itemize}
\item 
The first scenario involves NP contributions to the Wilson coefficients of the electroweak penguin operators of the SM, either $Q_8$ and $Q_{10}$, where the scale of NP is found to be around 4\,TeV, or $Q_7$ and $Q_9$, requiring a NP scale of approximately 1.5\,TeV. The required fine-tuning is of $\mathcal{O}(10^{-2})$.
\item 
The second scenario involves NP contributions to the Wilson coefficients of the operators $Q_1$ and $Q_2$ of the SM. Here the effective NP scale is required to be around 2.3\,TeV. This scenario can be ruled out based on the observation that the predictions for the branching ratios of the decays $B^+\to\pi^0 K^+$ and $B_d\to\pi^0 K^0$ deviate by almost two orders of magnitude from the experimental values.
\item 
The third scenario involves flavor-specific NP contributions in a more general basis of four-quark operators. Equations~\eqref{Rss linear approx} and \eqref{Rud linear approx} exhibit the strong correlation of the NP contributions to $R_{KK}^{ss}$ and $R_{\pi K}^{ud}$. This leads to much smaller NP contributions to the ratio $R_{KK}^{ss}/R_{\pi K}^{ud}$, see \eqref{Rss/Rud linear approx}, thereby excluding scenarios in which the NP contributions are smaller than the contributions of the SM. This fact has been exemplified by a fit allowing for NP contributions to the two leading operators, $Q_{6,u}^s$ and $Q_{6,d}^s$. The fit is fine-tuned, and the best fit value for $C_{6,d}^s$ is unsuppressed with respect to the SM, corresponding to an effective NP scale of approximately 4.5\,TeV. In this scenario, the CP-averaged branching fractions for the various $B_s\to K\bar K$ and $B\to\pi K$ decays are reproduced within the theoretical uncertainties. However, the predictions for the CP asymmetries in $B_s\to K^+ K^-$ decay show some tensions with experiment.
\end{itemize}

The specific numerical values that we extracted for the NP Wilson coefficients depend on hadronic parameters that suffer from rather large theoretical uncertainties. Our main aim, however, was not to perform precise calculations (which for hadronic $B$-meson decay is currently still not possible), but rather to make the point that a NP explanation of the three puzzles pointed out in \cite{Amhis:2022hpm} requires substantial fine-tuning and is rather implausible. A resolution of the puzzles related to the low value of ${\rm Br}(B_s\to K^0\bar K^0)$ must await further information from experiments.

\section*{Acknowledgements}

The work of YG is supported in part by the NSF grant PHY1316222 and by the United States--Israel Binational Science Foundation (grant number 2018257). YN is the Amos de-Shalit chair of theoretical physics, and is supported by grants from the Israel Science Foundation (grant number 1124/20), the United States--Israel Binational Science Foundation (grant number 2018257), the Minerva Foundation (with funding from the Federal Ministry for Education and Research), and the Yeda-Sela Center for Basic Research. The research of MN has received funding from the Cluster of Excellence PRISMA$^+$ (EXC 2118/1) within the German Excellence Strategy (Project-ID 390831469) and from the European Research Council (ERC) under the European Union’s Horizon 2022 Research and Innovation Program (ERC Advanced Grant agreement No.101097780, EFT4jets).

\begin{appendix}

\section{Details on the QCD factorization approach}

The vertex corrections encountered in \eqref{eq:treeeffects} are given by
\begin{equation}
\begin{aligned}
   V_{3,4}(M_2) 
   &= \int_0^1\!dx\,\Phi_{M_2}(x) \left[ 12\spac\ln\frac{m_b}{\mu} - 18 + g(x) \right] , \\
   V_5(M_2)
   &= \int_0^1\!dx\,\Phi_{M_2}(x) \left[ - 12\spac\ln\frac{m_b}{\mu} + 6 - g(1-x) \right] , \\
   V_6(M_2)
   &= \int_0^1\!dx\,\Phi_{m_2}(x)\,\big[ - 6 + h(x) \big] \,,
\end{aligned}
\end{equation}
with functions $g(x)$ and $h(x)$ given in equation (38) of \cite{Beneke:2003zv}. The hard spectator contributions in \eqref{eq:treeeffects} take the form
\begin{align}\label{eq:Hkernels}
   H_{3,4}(M_1 M_2) 
   &= \frac{B_{M_1 M_2}}{A_{M_1 M_2}}\,\frac{m_{B_q}}{\lambda_{B_q}} 
    \int_0^1\!dx \int_0^1\!dy \left[ \frac{\Phi_{M_2}(x)\,\Phi_{M_1}(y)}{(1-x)\spac(1-y)} 
    + r_\chi^{M_1}\,\frac{\Phi_{M_2}(x)\,\Phi_{m_1}(y)}{x\spac(1-y)} \right] , \notag\\
   H_5(M_1 M_2) 
   &= - \frac{B_{M_1 M_2}}{A_{M_1 M_2}}\,\frac{m_{B_q}}{\lambda_{B_q}} 
    \int_0^1\!dx \int_0^1\!dy \left[ \frac{\Phi_{M_2}(x)\,\Phi_{M_1}(y)}{x\spac(1-y)} 
    + r_\chi^{M_1}\,\frac{\Phi_{M_2}(x)\,\Phi_{m_1}(y)}{(1-x)\spac(1-y)} \right] , \notag\\
   H_6(M_1 M_2)
   &= 0 \,.
\end{align}
Here $\Phi_{M_i}(x)$ denotes the leading-twist LCDA of meson $M_i$, while $\Phi_{m_i}(x)$ refers to its twist-3 two-particle LCDA. We approximate the leading-twist LCDAs of pions and kaons in terms of their Gegenbauer expansions, truncated after the second Gegenbauer moment. For the twist-3 LCDAs we use the asymptotic forms in the Wandzura--Wilczek approximation, as discussed in \cite{Beneke:2003zv}. The parameter $\lambda_B$ is the first inverse moment of the leading-twist LCDA of the $B$ meson. The penguin coefficients in \eqref{eq:penguineffects} are given by
\begin{align}
   P_{4,q}(M_2) 
   &= \frac{C_F\spac\alpha_s}{4\pi\spac N_c} \left\{ C_{3,q}^D \left[ \frac43\spac\ln\frac{m_b}{\mu}
    + \frac23 - G_{M_2}(s_q) \right] \left( \delta_{qD} + \delta_{qb} \right) \right. \notag\\
   &\hspace{1.9cm} \left. + \big( C_{4,q}^D + C_{6,q}^D \big) \left[ \frac43\spac\ln\frac{m_b}{\mu}
    - G_{M_2}(s_q) \right] 
    - \left( \delta_{qD} + \delta_{qb} \right) C_{5,q}^D \int_0^1\!dx\,\frac{\Phi_{M_2}(x)}{1-x} \right\} , \notag\\
   P_{6,q}(M_2) 
   &= \frac{C_F\spac\alpha_s}{4\pi\spac N_c} \left\{ C_{3,q}^D \left[ \frac43\spac\ln\frac{m_b}{\mu}
    + \frac23 - \widehat G_{M_2}(s_q) \right] \left( \delta_{qD} + \delta_{qb} \right) \right. \notag\\
   &\hspace{1.9cm} \left. + \big( C_{4,q}^D + C_{6,q}^D \big) \left[ \frac43\spac\ln\frac{m_b}{\mu}
    - \widehat G_{M_2}(s_q) \right] 
    - \left( \delta_{qD} + \delta_{qb} \right) C_{5,q}^D \right\} ,
\end{align}
where $s_q=m_q^2/m_b^2$, and the functions $G_M(s)$ and $\widehat G_M(s)$ have been defined in equations (40) and (43) of \cite{Beneke:2003zv}, respectively. We treat the three light quark flavors $u$, $d$, $s$ as being massless but keep the masses of the charm and bottom quarks. The terms proportional to $C_{5,D}^D$ and $C_{5,b}^D$ in these expressions result from the contributions of these Wilson coefficients to the coefficient $C_{8g}^{\rm eff}$ of the chromo-magnetic dipole operator in the effective weak Hamiltonian (see e.g.\ \cite{Beneke:2001ev}).

\end{appendix}

\end{document}